\documentclass[traditabstract]{aa}

\usepackage[intlimits,sumlimits]{amsmath}

\usepackage{fixltx2e}
\usepackage{amssymb}
\usepackage{epsfig,amsmath,amssymb,amsfonts,mathrsfs,latexsym,graphicx}

\usepackage{natbib}

\usepackage{ textcomp }

\usepackage{amsmath}
\usepackage{amssymb}
\usepackage[english]{babel}
\usepackage{chngcntr}
\include{journals}
\newcommand{\beqa}{\begin{eqnarray}} 

\newcommand{\eeqa}{\end{eqnarray}}

\newcommand{\bsub}{\begin{subequations}}
\newcommand{\esub}{\end{subequations}}
\newcommand{\beal}{\begin{align}}
\newcommand{\ealn}{\end{align}}
\newcommand{\lm}{$\mathrm{L_{max}}$~}
\newcommand{\lmn}{$L_{max}$}
\newcommand{\tb}{$t_B$(max)\,}
\newcommand{\tbn}{$t_B$(max)}
\newcommand{\Mni}{$\rm M_{^{56}Ni}$~}
\newcommand{\Mnin}{$\rm M_{^{56}Ni}$}

\newcommand{\ejm}{M$_{ej}$\,}
\newcommand{\ejmn}{M$_{ej}$}
\newcommand{\dm}{$\Delta m_{15}(B)$\,}
\newcommand{\mc}{$\rm M_{Ch}$\,}
\newcommand{\Nif}{$
\rm ^{56}Ni$\,}

\newcommand{\s}{\rm M$_{\sun}$~}
\newcommand{\sfa}{{\em SN~Ia-faint}}
\newcommand{\sn}{\rm M$_{\sun}$}
\newcommand{\sbv}{$s_{BV}$\,}
\newcommand{\egs}{$\times$ 10$^{43}$~erg~s$^{-1}$}
\newcommand{\kms}{$\mathrm{km\,s^{-1}}$}

\def\gsim{\mathrel{\rlap{\lower 4pt \hbox{\hskip 1pt $\sim$}}\raise 1pt \hbox {$>$}}}
\def\lsim{\mathrel{\rlap{\lower 4pt \hbox{\hskip 1pt $\sim$}}\raise 1pt \hbox {$<$}}}

\counterwithout{table}{section}

\begin{document}
\title{Two classes of fast-declining type Ia supernovae}
\titlerunning{Fast declining SNe~Ia}
\authorrunning{S. Dhawan et al.}
\author{\textbf{Suhail Dhawan\inst{1,2,3}
 	\and B. Leibundgut\inst{1,2}
    \and J. Spyromilio\inst{1}
    \and S. Blondin\inst{4}
    }}

\institute{European Southern Observatory, Karl-Schwarzschild-Strasse 2, D-85748 Garching bei M\"unchen, Germany \\
\email{sdhawan@eso.org}
\and  Excellence Cluster Universe, Technische Universit\"at M\"unchen,
Boltzmannstrasse 2, D-85748, Garching, Germany\\
\and Physik Department, Technische Universit\"at M\"unchen, James-Franck-Strasse 1, D-85748 Garching bei M\"unchen\\
\and Aix Marseille Universit\'e, CNRS, LAM, Laboratoire d'Astrophysique de Marseille,  Marseille, France\\
} 

\date{Received; accepted }

\offprints{S. Dhawan}

\abstract{
Fast-declining Type Ia supernovae (SN\,Ia) separate into two categories based on their bolometric and near-infrared (NIR) properties. The peak bolometric luminosity (\lm), the phase of the first maximum relative to the optical, the NIR peak luminosity and the occurrence of a second maximum in the NIR distinguish a group of very faint SN\,Ia. 
Fast-declining supernovae show a large range of peak bolometric luminosities ($L_{max}$ differing by up to a factor of $\sim$ 8). All fast-declining SN\,Ia with $\mathrm{L_{max}} < 0.3$ \egs \ are spectroscopically classified as 91bg-like and show only a single NIR peak. SNe with $\mathrm{L_{max}} > 0.5$ \egs \ appear to smoothly connect to normal SN\,Ia. The total ejecta mass (\ejmn) values for SNe with enough late time data are $\lesssim$1\,$M_{\odot}$, indicating a sub-Chandrasekhar mass progenitor for these SNe. 

}

\maketitle
\section{Introduction}
\label{sec-intro}

Type\,Ia supernovae have long  been linked with the explosion of a C/O white dwarf \citep{Hoyle1960}. Ignition of the white dwarf can lead to fusion of (the) C/O to $^{56}$Ni releasing enough energy to unbind the progenitor and through the deposition of the energy released in the radioactive decay of $^{56}$Ni to $^{56}$Co, and on to $^{56}$Fe, power the electromagnetic display of the supernova. This scenario is extremely robust and is supported both by theoretical studies \citep{Hillebrandt2000} and observations over many decades, although the exact white dwarf mass and ignition scenario remain the subject of extensive debate. 

In the past two decades the study of supernovae has been blessed with an great increase in high quality data through a series of systematic surveys for transients and extended temporal and wavelength coverage. Dedicated supernova searches have discovered several SN\,Ia with unusual photometric and spectroscopic properties.  Some peculiar SN\,Ia exhibit fast optical post-peak declines and a deep trough-like feature at $\sim$ 4200 \AA\, in their maximum light spectra, attributed to Ti\,II. The prototypical example of this class is SN\,1991bg \citep{Filippenko1992b,Leibundgut1993,Mazzali1997}. \citet{Li2011} showed that SN\,1991bg-like events comprise a large fraction (15-20$\%$) of the SN\,Ia population in a volume-limited sample and appear distinct from normal SN\,Ia in their width-luminosity relationship.  SN\,1991bg-like events have been shown to prefer elliptical and lenticular galaxies \citep{Howell2001}.

Based on multi-epoch spectra and multi-band optical light curves of a sample of fast-declining, SN\,1991bg-like SN\,Ia, \citet{Tauben2008}  suggested that this class may have a different physical origin to normal SN\,Ia, although the possibility that they are a low-luminosity, fast-declining extension  of normal SN\,Ia cannot be excluded. These SNe show markedly different optical colour evolution and low \Nif mass values as calculated from UBVRI pseudo-bolometric light curves. Supernovae with intermediate properties between normal and sub-luminous SN\,Ia would lend support to the latter hypothesis \citep{Garnavich2004}.

The optical width-luminosity relation for SN\,Ia \citep{Phillips1999, Burns2011} shows a notable break for fast-declining objects (\dm \textgreater 1.6). Fast-declining SN\,Ia are fainter given their \dm\, assuming a linear relation,  possibly due to the inability of \dm to properly characterise fast-declining SNe since their light curves settle onto a linear magnitude decline at approximately 15\,days past B maximum. \citet{Burns2014} proposed a different ordering parameter, \sbv, to improve the treatment of fast-declining objects. $s_{BV}$ is defined as the epoch at which the $(B-V)$ colour curve is at its  maximum value, divided by 30\,d. Using this metric the fast-declining SNe appear less distinct and more as a continuous tail of the distribution of normal SN\,Ia.

In the near infrared SN\,Ia are remarkably uniform around maximum \citep{Elias1981,Meikle2000,Krisciunas2004, Folatelli2010, Dhawan2015}. While a majority of SN\,Ia show a homogeneous behaviour around the maximum, there are some clear outliers.
\citet{Garnavich2004} reported that the 91bg-like SN\,1999by was fainter in the NIR ($JHK$ filters) than the average derived for normal SN\,Ia in \citet{Krisciunas2004}. Subsequent studies found a bi-modality in the NIR light curve properties of fast-declining SN\,Ia \citep[e.g.][]{Krisciunas2009,Folatelli2010,Kattner2012,Phillips2012}. Events whose NIR primary maxima occur after $B$-band maximum (\tbn) are sub-luminous in all bands compared to normal SN\,Ia. These sub-luminous SN\,Ia also tend to lack or have very weak second maxima in their NIR light curves. However, objects that peak in the NIR \emph{before} \tb have NIR absolute magnitudes comparable to normal SN\,Ia and show  prominent (albeit, early) second maxima. Following these results, \citet{Hsiao2015} proposed the definition of `transitional' SNe as fast-declining SN\,Ia with an NIR maximum before \tb.

In this paper we analyse the NIR and bolometric properties of fast-declining SN~Ia to determine whether they are an extension of normal SN~Ia or a distinct subclass. In section\,\ref{sec-data} we describe our sample and in section\,\ref{sec-lmsbv} we show that fast-declining SN\,Ia are found in two distinct groups. In sections~\ref{ssec-max} and \ref{ssec-ni_ej} we examine other distinguishing characteristics of the groups. The discussion and conclusions are presented in sections~\ref{sec-disc}.

\section{Data} 
\label{sec-data}

We compiled a sample of fast-declining SN\,Ia with $\Delta m_{15}>1.6$ from the literature. We do not include objects similar to 2002cx \citep[dubbed `Type Iax' supernovae][]{Foley2013}. Some of Iax SNe are fast decliners (e.g. SN\,2002cx, \citealt{Li2003};  SN\,2005hk, \citealt{Jha2006, Phillips2007}; SN\,2008ha, \citealt{Foley2009}) and could have been included in our sample, but the evidence for them being different kinds of explosions is mounting  \citep[e.g][]{Li2003,Jha2006}. We discuss some of the SN\,Iax features in the conclusions. 

Most of our data is compiled from the Carnegie Supernova Project \citep[CSP;][]{Contreras2010,Stritzinger2011} augmented by the CfA supernova survey on PAIRITEL \citep{Wood-Vasey2008,Friedman2014}. To these objects we add SN1999by \citep{Garnavich2004} and iPTF13ebh \citep{Hsiao2015}. The objects in  our sample along with the sources of the data are presented in Table \ref{tab:samp}.

Our sample has 15 SNe. Ten of these SNe are spectroscopically classified as 91bg-like \citep{Garnavich2004, Folatelli2013}. Seven SNe in our sample show a pronounced NIR second maximum, three of which are spectroscopically 91bg-like (2006gt, 2007ba and 2008R). 

 For SNe with z \textgreater 0.01, we use luminosity distances with H$_0$ = 70 km s$^{-1}$Mpc$^{-1}$, $\Omega_m$=0.27 and $\Omega_\Lambda$ = 0.73. For nearby SNe with z \textless 0.01 we use independent distances to the host galaxy from the literature. A summary of the methods used for the distances and the references is provided in Table~\ref{tab:samp}

For SNe observed by the CSP, we have used published values of \sbv. For other SNe we calculated the \sbv from SNooPY \citep{Burns2011} fits to the data.

\begin{table*}
\caption{SN sample used in this analysis.}
\centering
\scalebox{.71}{\begin{tabular}{lllccccccccc}
\hline\hline
SN & $t_B(max)$ & $\Delta m_{15}$(B) & $s_{BV}$ &  \sbv Reference & $t_2$(Y) \tablefootmark{1} & $t_2$(J) & $\mu$ & Distance Method \tablefootmark{**} &  Distance Reference & Data Reference \\
 &  \multicolumn{1}{c}{(MJD; d)} &  \multicolumn{1}{c}{(mag)} & & \multicolumn{1}{c}{(d)} & \multicolumn{1}{c}{(d)} & \multicolumn{1}{c}{(mag)}   & \\
\hline
{\em SN1999by} \tablefootmark{*} \tablefootmark{2} & 51308.3 & 1.93 & 0.46 &  This paper &  N/A & N/A & 30.82 ($\pm$ 0.15)  & TF & T13 & H02,G04\\
SN2003gs & 52848.3	&	1.83	& 0.49 &  This paper & $\cdots$ & 15.3 ($\pm$ 0.7) & 	31.65 ($\pm$ 0.28) & SBF & T01 & K09\\
{\em SN2005bl}$^{*}$		& 53481.6	& 1.80	& 0.39 &  B14 & N/A & N/A  & 35.14 ($\pm$ 0.09) & LD & &  WV08, F15	\\
{\em SN2005ke}$^{*}$ & 53698.6  & 1.78 & 0.41 &   B14	& N/A & $\cdots$ & 31.84 ($\pm$ 0.08) &	SBF & T01 & WV08, C10, F15\\
SN2006gt$^{*}$	& 54003.1	&	1.66 &	0.56 &   B14 & $\cdots$ & 20.2 ($\pm$ 1.2) & 36.43 ($\pm$ 0.05) & LD & &  C10			\\
{\em SN2006mr}$^{*}$ & 54050.2 	& 1.84 &	0.26 &  B14 & N/A & N/A & 31.15 ($\pm$ 0.23) & SBF & A01 & C10 \\
{\em SN2007N}$^{*}$ & 54124.3 	&1.79	&	0.29 &  B14 & N/A & N/A & 33.91 ($\pm$ 0.16) & LD & & S11 \\
{\em SN2007ax}$^{*}$ & 54187.5	&1.86 	& 0.36	& B14 & N/A & N/A & 32.20 ($\pm$ 0.14) & TF & T09 & S11 \\
SN2007ba$^{*}$ & 54196.2 	 & 1.88	&	0.54 & B14 & 20.0 ($\pm$ 0.4) & $\cdots$ & 36.18 ($\pm$ 0.05) & LD & &  S11	\\
SN2007on & 54421.1 	& 1.90	& 0.57	&  B14  & 18.7 ($\pm$ 0.4) & 18.2 ($\pm$ 0.1) & 31.45 ($\pm$ 0.08) & SBF  & J03 &  S11 \\
SN2008R$^{*}$	 &	54494.3 & 1.85	& 0.59   &  B14 &15.5 ($\pm$ 0.7) & 14.1 ($\pm$ 0.7)& 33.73 ($\pm$ 0.16) & LD & &  S11	\\
SN2008hs &	54812.1 & 1.83	& 0.60 &   This paper & $\cdots$ & 14.0 ($\pm$ 1.0) & 34.28  ($\pm$ 0.13) & LD & & F15	\\
{\em SN2009F}$^{*}$ & 54841.8 & 1.97 & 0.33 &  B14 & N/A & N/A & 33.73 ($\pm$ 0.16) & LD &  & S11	\\
SN2010Y	& 55247.5 &1.73 & 0.61  &  This paper & $\cdots$ & $\cdots$ & 33.44 ($\pm$ 0.20) & LD & &	F15	\\
iPTF13ebh & 56622.9	& 1.79	&	0.63	&  H15 & 19.4 ($\pm$ 0.2) & 17.2 ($\pm$ 1.5) &	33.63 ($\pm$ 0.16)	& LD & &	H15 \\
\hline
\end{tabular}}
\label{tab:samp}
\tablefoot{\tablefoottext{1}{SN\,Ia with only one maximum are marked as `N/A'. Ellipses indicate insufficient data to determine a second maximum.}\\
\tablefoottext{2}{SN\,Ia in the {\em SN\,Ia-faint} group are shown in {\em italics} (see text)}\\
\tablefoottext{*}{Spectroscopically classified as SN\,1991bg-like}\\
\tablefoottext{**}{Methods for distances to the SN hosts are as follows: LD: luminosity distance (using parameters detailed in the text)}, TF: Tully-Fisher relation, SBF: surface brightness fluctuation. Note that 0.16 mag \citep{Jensen2003} is subtracted from SBF distances from \citet{Tonry2001} to put them on the same scale as \citet{Freedman2001}. Note that objects that do not have a luminosity distance presented here are not in the Hubble flow i.e. have z \textless 0.01.}

\tablebib{A01: \citet{Ajhar2001}, T01: \citet{Tonry2001}, H02: \citet{Hoeflich2002}, J03: \citet{Jensen2003}, G04: \citet{Garnavich2004}, K09: \citet{Krisciunas2009}, WV08: \citet{Wood-Vasey2008}, C10: \citet{Contreras2010}, S11: \citet{Stritzinger2011},  T09: \citet{Tully2009}, T13: \citet{Tully2013}, F15: \citet{Friedman2014},  H15: \citet{Hsiao2015}}

\end{table*}

\section{Luminosity vs colour-stretch: Evidence for two classes of fast-declining SNe}
\label{sec-lmsbv}

As has been discussed in \citet{Dhawan2015} the infrared spectral region (JHK) is a significant contributor to the bolometric luminosity of SN\,Ia. 
We calculate the pseudo-bolometric light curve by integrating over a $u \rightarrow H$ (UVOIR) SED based on monochromatic fluxes derived using the transmission curves for each survey \citep[see][for a detailed explanation of the method]{Contardo2000}.  The light curves are corrected for host galaxy and Milky Way extinction
We determine the  absolute UVOIR peak luminosity (\lmn) by fitting a cubic spline to the constructed pseudo-bolometric light curve. 

\begin{figure*}
\centering
\includegraphics[width=.9\textwidth]{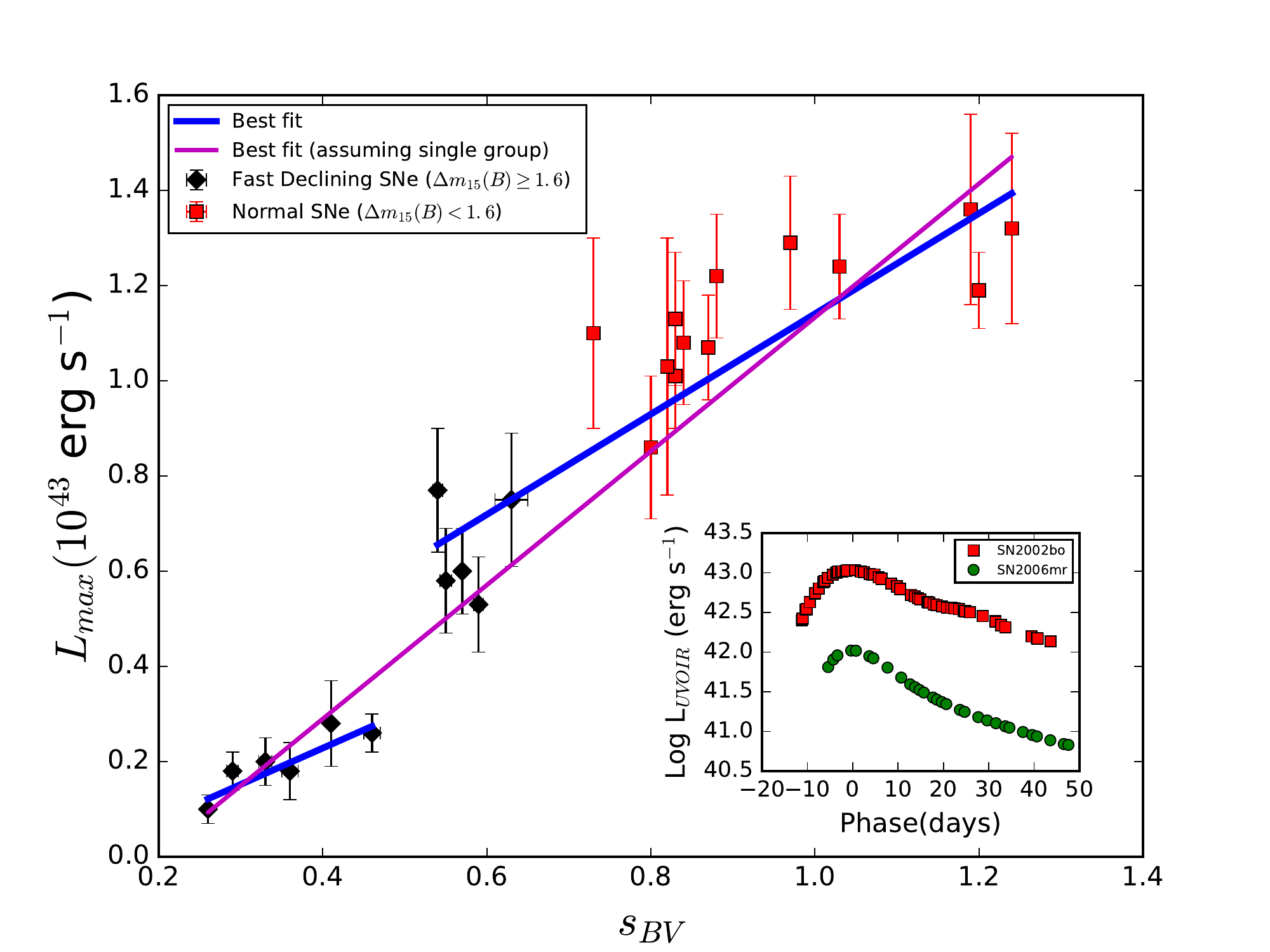}
\caption{$L_{max}$ versus $s_{BV}$ for normal SN\,Ia (red) and fast-declining SN\,Ia (black). The best fit linear relations for the faint sub-group of the fast-declining SN\,Ia, the normal SNe, as well as the best fit assuming that all SNe belong to the same group are plotted as solid lines.  The \lm values for the normal SN~Ia were calculated in \citet{Dhawan2015b} and the \sbv values are from \citet{Burns2014}. \emph{Inset}: The $u \rightarrow H$ pseudo-bolometric light curve for SN~2006mr (\emph{green}), the faintest SN in the sample is plotted in comparison with the normal SN~2002bo \citep[\emph{red};][]{Benetti2004}. From the bolometric light curves it is clear that SN~2006mr has a faster post-maximum evolution and settles earlier onto the exponential tail than SN~2002bo which was well described by a \mc\, delayed detonation model  \citep{Benetti2004, Blondin2015}.  The reduced $\chi^2$ for the single population is 1.90 whereas for the two subclasses is 1.03}
\label{fig:ni-sbv}
\end{figure*}

The \sbv versus \lm relationship (Figure~\ref{fig:ni-sbv}) exhibits two distinct groups among the fast declining SN\,Ia. One group extends the trend of normal Ia supernovae with lower luminosity having an earlier \sbv while a second group is detached and appears to follow a different relation. The relation for the faint subgroup has a significantly different slope ($\sim 2\sigma$ level; Table~\ref{tab:slope}).  We note that the error in the slope for fitting all SNe as single group is lower than the two separate subgroups. However, this is because the total sample size is greater than the comparatively smaller subsamples of the two separate groups. 

 A simple $\chi^2$/DoF analysis shows that a fit to two subclasses (fitting two slopes simultaneously) is favoured (reduced $\chi^2=1.03$) compared to a single line fit (reduced $\chi^2=1.90$). We also apply the hypothesis testing technique of comparing the logarithm of the Bayesian Evidence \citep[$lnZ$; see][for details about ]{Skilling2004}. $Z$ is the integral of the likelihood over the prior region

\begin{equation}
Z = \int L \pi d \theta
\label{eq:evidence}
\end{equation}

where $L$ is the likelihood, $\pi$ is the prior and $\theta$ is the set of the parameters. 
We calculate $lnZ$  using a multi-modal nested sampling algorithm, \texttt{MultiNest} \citep{Feroz2013}. The $\Delta ln Z$ for the single population relative to the two subclasses is $\sim$ -6.97 suggesting a strong preference for the two subclasses over the single population \citep[][suggest that $\Delta ln Z$  \textless -5 is strong evidence for the alternate model over null hypothesis]{Trotta2008}.

This is an intriguing result suggesting that there are two separate populations of fast-declining SN\,Ia which we refer to as {\em SN\,Ia-faint} here, as opposed the group that appears to join the normal SN\,Ia. The fainter group consists of SNe\,1999by, 2005ke, 2006mr, 2007N, 2007ax and 2009F. 

\begin{table}
\centering
\caption{Slope and intercept for the \lm-\sbv relation}
\label{tab:slope}
\scalebox{.85}{\begin{tabular}{lccr}
\hline\hline
Sample & Slope & Intercept &\\
\hline
{\em SN\,Ia-faint}	& 0.76 ($\pm$ 0.17) 	&	-0.08 ($\pm$ 0.06)	&	\\
Normal SN\,Ia	& 1.24 ($\pm$ 0.14)		&   -0.05 ($\pm$ 0.11)	&	\\
Complete &  1.40 ($\pm$ 0.08) &  -0.27 ($\pm$ 0.04) & \\
\hline
\end{tabular}}
\end{table}

We note that \citet[their Fig.~10]{Burns2014} find a the relation between the pseudo equivalent width of the Si II 5972 \AA\ line and $s_{BV}$, which has a more complicated form than a simple linear relation for SNe with \sbv \textless 0.5. This would be further evidence that the {\em SN\,Ia-faint} group of fast-declining SN\,Ia are a separate population. 

\section{Characterising fast-declining, low-luminosity SN\,Ia}
There appear to be at least three distinguishing characteristics of the {\em SN\,Ia-faint} subgroup. In addition to their extreme low luminosity, they appear to reach the NIR peak at a later stage compared to optical wavelengths and do not show a second maximum in their near-infrared light curves. 

\subsection{Phase of the first NIR maximum}
\begin{center}
\begin{figure}
\includegraphics[width=.5\textwidth]{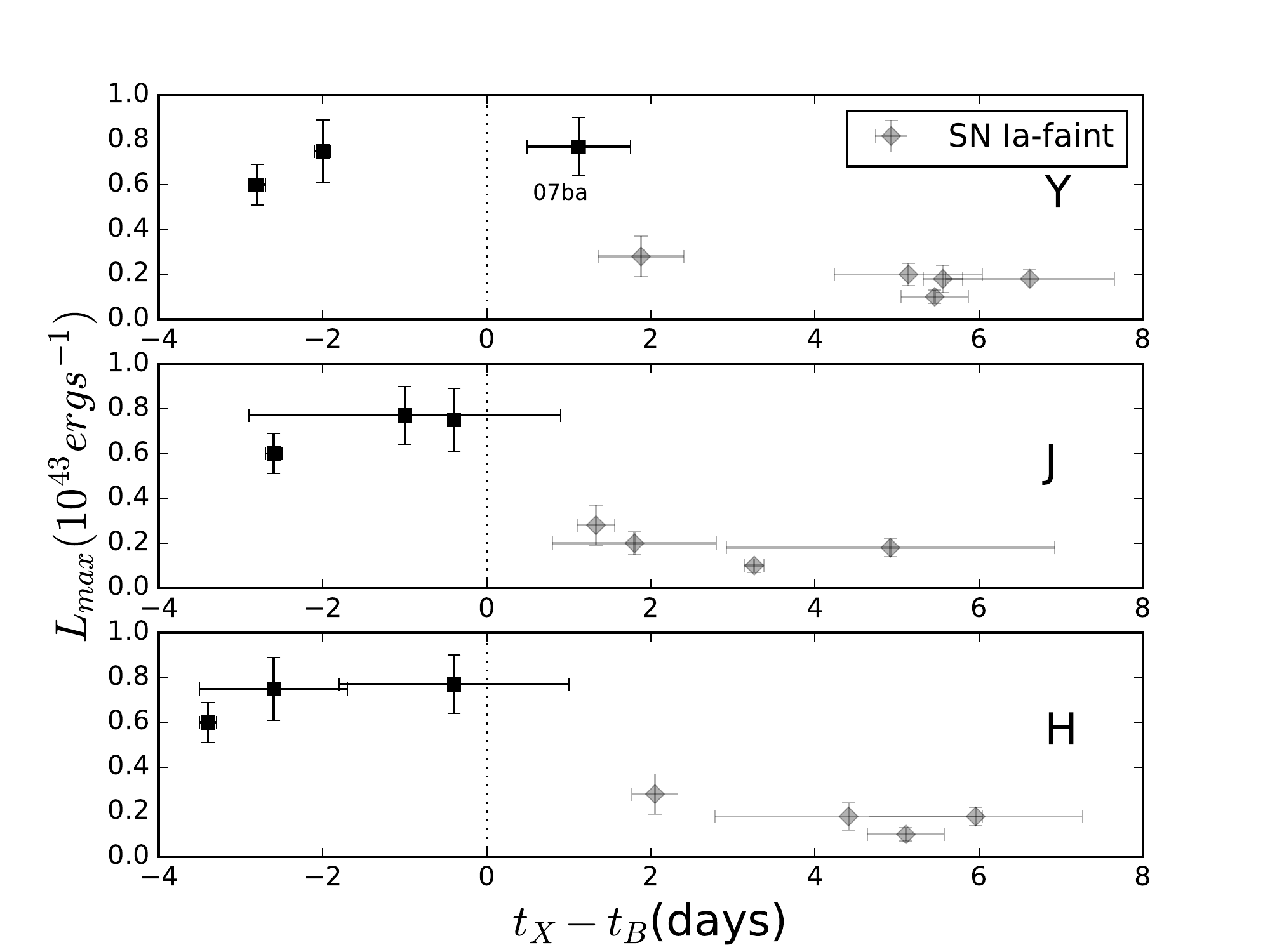}
\caption{The pseudo-bolometric peak luminosity versus the timing of the NIR ($YJH$ filters) maximum for SNe in our sample. SNe that peak \emph{earlier} in the NIR (\emph{squares}) than the optical also appear to be an extension to the normal SN\,Ia whereas all SNe with an NIR peak \emph{after} the optical (\emph{diamonds}) appear to be a distinct population in Figure~\ref{fig:ni-sbv}. We note that in the $Y$-band there is one exception, SN\,2007ba which has a high bolometric peak luminosity but a $Y$-band first maximum shortly \emph{after} \tb}
\label{fig:max}
\end{figure}
\end{center}

As originally pointed out by \citet{Contardo2000} and extended by future studies  \citep{Krisciunas2004, Kattner2012, Dhawan2015} regular SN\,Ia reach the first maximum in the infrared several days earlier than in the optical bands. The physical reason for this is not entirely obvious, but could be due to the rapid shift of the SED to the blue as the spectrum-formation region heats up \citep{Blondin2015}. Several low-luminosity SN\,Ia reach their NIR peaks after the optical maximum \citep{Krisciunas2004, Kattner2012}. We have investigated this for our sample. The magnitude and epoch of the NIR maximum for our sample is reported in Table~\ref{tab:mt} and displayed against the bolometric peak luminosity in Fig.~\ref{fig:max}. The separation between the two groups is a function of wavelength and appears to increase from $Y$ to $H$, with the effect being evident in $J$ and $H$ and less visible in $Y$. Similar to Fig.~\ref{fig:ni-sbv} the luminosity appears to be the distinguishing property, while the timing of the NIR maximum wrt \tb\, seems more like a continuous distribution. Nevertheless, the objects in our {\em SN\,Ia-faint} are separated from the normal SN\,Ia. We also note that these objects are also sub-luminous in the NIR filters, whereas SNe which peak before \tb are \emph{not sub-luminous}. 

\begin{table*}
\centering
\caption{Epoch of maximum (with respect to \tb) and peak magnitude in $YJHK$ filters}
\begin{tabular}{cccccccc}
\hline\hline
SN\tablefootmark{1} & $t_Y$ & $t_J$ & $t_H$ &  $M_Y$ & $M_J$ & $M_H$  \\
& (d) & (d) & (d) & (mag) & (mag) & (mag)\\
\hline
{\em 1999by} & \ldots & \ldots & 2.98 ($\pm$ 0.63) & \ldots & \ldots & -18.33 ($\pm$ 0.19) \\
{\em 2005bl} & \ldots & 1.12 ($\pm$ 1.09) & \ldots & \ldots & -17.96 ($\pm$ 0.13) & \ldots \\
{\em 2005ke} & 1.88 ($\pm$ 0.52) & 1.33 ($\pm$ 0.23) & 2.05 ($\pm$ 0.28) & -17.42 ($\pm$ 0.08) & -17.45 ($\pm$ 0.08) & -17.50 ($\pm$ 0.08) \\
{\em 2006mr} & 5.46 ($\pm$ 0.41) & 3.26 ($\pm$ 0.12) & 5.11 ($\pm$ 0.47) & -17.17 ($\pm$ 0.23) & -17.17 ($\pm$ 0.23) & -17.27 ($\pm$ 0.23) \\
{\em 2007N} & 6.62 ($\pm$ 1.03) & 4.92 ($\pm$ 2.00) & 5.96 ($\pm$ 1.30) & -17.48 ($\pm$ 0.16) & -17.48 ($\pm$ 0.18) & -17.65 ($\pm$ 0.21) \\
{\em 2007ax} & 5.56 ($\pm$ 0.24) & \ldots & 4.41 ($\pm$ 1.63) & -17.01 ($\pm$ 0.14) & \ldots & -17.00 ($\pm$ 0.15) \\
2007ba & 1.12 ($\pm$ 0.63) & -1.05 ($\pm$ 1.9) & -0.42 ($\pm$ 1.40) & -18.65 ($\pm$ 0.06) & -18.56 ($\pm$ 0.31) & -18.64 ($\pm$ 0.11) \\
2007on & -2.88 ($\pm$ 0.10) & -2.67 ($\pm$ 0.10) & -3.49 ($\pm$ 0.10) & -18.28 ($\pm$ 0.19) & -18.37 ($\pm$ 0.19) & -18.18 ($\pm$ 0.19) \\
2008hs & \ldots & -2.77 ($\pm$ 0.63) & -3.21 ($\pm$ 1.69) & \ldots & -17.96 ($\pm$ 0.15) & -17.82 ($\pm$ 0.15) \\
{\em 2009F} & 5.14 ($\pm$ 0.90) & 1.80 ($\pm$ 1.00) & \ldots & -17.64 ($\pm$ 0.19) & -17.57 ($\pm$ 0.17) & \ldots \\
2010Y & \ldots & -1.88 ($\pm$ 1.70) & \ldots & \ldots & -18.21 ($\pm$ 0.26) & \ldots \\
iPTF13ebh & -2.02 ($\pm$ 0.10) & -0.48 ($\pm$ 0.09) & -2.62 ($\pm$ 0.91) & -18.57 ($\pm$ 0.16) & -18.58 ($\pm$ 0.16) & -18.46 ($\pm$ 0.16) \\
\hline
\end{tabular}
\label{tab:mt}
\tablefoot{\tablefoottext{1}{SN\,Ia in the {\em SN\,Ia-faint} group are shown in {\em italics}}}
\end{table*}

\subsection{Lack of NIR second maximum}
\label{ssec-max}

 A further characteristic property of {\em SN\,Ia-faint} is a lack of a second infrared maximum. Table~\ref{tab:samp} indicates the phase of the second maximum when it could be measured. Objects without a second maximum are labelled 'N/A' and correspond to the low-luminosity objects. There is a clear separation of the class of {\em SN\,Ia-faint} in this respect. We cannot confirm the UV to NIR (UVOIR) luminosity of SN\,2005bl independently, but the lack of a second maximum, the late phase of the first maximum and low \sbv, indicate that this object also belongs to the {\em SN\,Ia-faint} subgroup \footnote{Moreover, a low \Nif\, mass ($\sim$ 0.1 $M_{\odot}$) from UBVRI light curve calculations by \citet{Tauben2008} lends further evidence to its classification as \emph{SN~Ia-faint}}.

\subsection{Low \Nif and ejecta mass}
\label{ssec-ni_ej}

We can further investigate the properties of fast-declining SN\,Ia by calculating the ejecta masses and production of \Nif. 

From our calculated \lmn, we estimate the \Nif mass using:

\begin{equation}
\label{eq:lm-ni}
L_{max} = 2.0 (\pm 0.3) \times 10^{43} \frac{\mathrm{M_{^{56}Ni}}}{\mathrm{M_{\odot}}} \mathrm{erg s^{-1}}.
\end{equation} 

This is a simple implementation of Arnett's rule \citep{Arnett1982, Arnett1985} for a rise time of 19\,days. Variations in Arnett's rule have been encapsulated in a parameter $\alpha$ \citep[see][]{Branch1992}. We use $\alpha=1$. \citet{Tauben2008} find that fast-declining SN\,Ia have shorter rise times (typically 13 -16\,d) which would imply lower \Nif masses by 40-15\,$\%$ for the same \lmn. The resulting \Nif masses for 13 and 19\,days rise times are reported in Table~\ref{tab:ni}. 

\begin{table*}
\label{tab:ni}
\caption{\Nif masses for fast-declining SN\,Ia with sufficient early time coverage to determine a peak luminosity}
\label{tab:ni}
\begin{tabular}{llcccccc}
\hline\hline
SN\tablefootmark{1} & \multicolumn{1}{c}{L$_{max}$}  & $M_{^{56}Ni}$ & \Mni  & E(B-V)$_{host}$ &  E(B-V)$_{MW}$ & R$_V$\tablefootmark{2}  & Reference for E(B-V)$_{host}$ \\
 &  \multicolumn{1}{c}{(10$^{43}$ erg s$^{-1}$)} & \multicolumn{1}{c}{$t_R$=13\,d; \s}   & \multicolumn{1}{c}{$t_R$=19\,d; \s} & \multicolumn{1}{c}{mag} & \multicolumn{1}{c}{mag} & \\
\hline
{\em 1999by} & 0.26	($\pm$ 0.04) 	& 0.10 ($\pm$ 0.02) & 0.13 ($\pm$ 0.03) &  0.020 $\pm$ 0.030  &  0.010 &  3.1 & G04\\
{\em 2005ke} & 0.28 ($\pm$ 0.09)	& 0.10 ($\pm$ 0.04) & 0.14 ($\pm$ 0.04) &  0.263 ($\pm$ 0.033) &  0.020  &  1.0 & B14\\
2006gt 		 & 	0.58 ($\pm$ 0.11) 	& 0.21 ($\pm$ 0.05) & 0.29 ($\pm$ 0.07) &  0.040 ($\pm$ 0.014) & 0.032 &	 3.1 & B14\\ 
{\em 2006mr} & 0.10 ($\pm$ 0.03) 	& 0.04 ($\pm$ 0.01) & 0.05 ($\pm$ 0.02)	&  0.089 ($\pm$ 0.039) & 0.018 &  2.9 & B14\\
{\em 2007N}  & 0.18 ($\pm$ 0.04)	& 0.07 ($\pm$ 0.02) & 0.09 ($\pm$ 0.03) &  0.350 ($\pm$ 0.052) &  0.034 &  1.7 & B14\\
{\em 2007ax} & 0.17 ($\pm$ 0.06) 	& 0.07 ($\pm$ 0.02) & 0.09 ($\pm$ 0.03) &  0.213 ($\pm$ 0.049) &  0.045 &  2.1 & B14\\
2007ba 		 & 0.77	($\pm$ 0.13) 	& 0.28 ($\pm$ 0.06) & 0.38 ($\pm$ 0.09) & 0.150 ($\pm$ 0.026) & 0.032 &  1.1 & B14\\
2007on 		 & 0.60 ($\pm$ 0.09) 	& 0.22 ($\pm$ 0.05) & 0.30 ($\pm$ 0.07) &  \textless 0.007 &  0.010 &  1.9 & B14\\
2008R		 & 0.53 ($\pm$ 0.10) 	& 0.20 ($\pm$ 0.05) & 0.27 ($\pm$ 0.07)	&  0.009 $\pm$ 0.013 &  0.062 &  3.1 & B14\\
{\em 2009F}  & 0.20 ($\pm$ 0.05)	& 0.07 ($\pm$ 0.02) & 0.10 ($\pm$ 0.03) &  0.108 ($\pm$ 0.047) &  0.089 &  1.0 & B14\\
iPTF13ebh 	 & 0.75 ($\pm$ 0.14)	& 0.28 ($\pm$ 0.07) & 0.38 ($\pm$ 0.10) &  0.050 ($\pm$ 0.020) &	 0.067  &  3.1 & H15\\
\hline
\end{tabular}
\tablefoot{\tablefoottext{1}{SN\,Ia in the {\em SN\,Ia-faint} group are shown in {\em italics}}}
\tablefoot{\tablefoottext{2}{R$_V$ values were calculated using SNooPY fits \citep{Burns2011,Burns2014}.}}
\tablebib{G04: \citet{Garnavich2004}, B14: \citet{Burns2014}, H15: \citet{Hsiao2015}}
\end{table*}

Unsurprisingly, the values of \Nif mass in Table ~\ref{tab:ni} are significantly lower than the averages derived for normal SN~Ia \citep[0.5 - 0.6 \sn; e.g.][]{Stritzinger2006a,Scalzo2014, Dhawan2015b}. The \Nif mass values we derive range from 0.05 - 0.38 \s indicating a significant diversity in the sample. A basic assumption in equation~\ref{eq:lm-ni} is that the ejecta mass is  the same for all SN~Ia when determining a nickel mass. If the {\em SN\,Ia-faint} have a  lower ejecta mass then the derived nickel masses estimates presented here are too large \citep{Pinto2000}. Of course equation~\ref{eq:lm-ni} does not apply if other processes than photon diffusion or a different energy source are at work in low-luminosity SN\,Ia. We allow the rise to vary between 13 and 19 days in our derivation of the \Nif\, mass. The results are presented in Tab.~\ref{tab:ni}. 

To calculate the ejecta mass we use  \citep[see][for a detailed derivation]{Jeffrey1999}: 
\begin{equation}
\label{eq:ejm}
M_{ej} = 1.38 \cdot \left(\frac{1/3}{q}\right) \cdot \left( \frac{v_e}{3000\, \mathrm{kms^{-1}}}\right)^2 \cdot \left(\frac{t_0}{36.80\,d}\right)^2 \mathrm{M_{\odot}}
\end{equation}
Equation\,\ref{eq:ejm} encapsulates the capture rate of $\gamma$-rays in an expanding spherical volume for a given distribution of the radioactive source. The e-folding velocity $v_e$ provides the scaling length for the expansion, $q$ is a qualitative description of the distribution of the material within the ejecta with 1/3 being a uniform distribution and higher values reflecting more centrally concentrated \Nif. We assume a constant $\gamma$ ray opacity of 0.025 cm$^2$ g$^{-1}$ \citep{Swartz1995}. 

The `fiducial' timescale ($t_0$) defined by \citet{Jeffrey1999} as a  parameter that governs the time-varying $\gamma$-ray optical depth behaviour of a supernova is the only `observable'. 

We determine $t_0$ by fitting the radioactive decay energy deposition to the late time (40---90\,days) bolometric light curve (see Equation~\ref{eq:dep}). As the UVOIR light curve is not truly bolometric there is an implicit assumption that the thermal infrared and the ultraviolet beyond the atmospheric cut off are not significant contributors. This assumption is supported by modelling that shows that the infrared catastrophe does not occur until much later and the line blanketing opacity in the UV remains high \citep{Blondin2015, Fransson2015}. The deposition function for re-processed photons is then given by the following equation:

\begin{equation}
\label{eq:dep}
E_{dep} = E_{Ni} + E_{Co, e^{+}} + [1 - exp(-\tau)] \cdot E_{Co, \gamma}
\end{equation}

Among the fast-declining SN\,Ia we have 5 objects for which we can determine both the maximum bolometric luminosity and the fiducial decay time scale, and hence, can derive both \ejm\ and \Mnin. As described above, the application of Arnett's rule implicitly assumes an ejecta mass through its impact on the diffusion timescale. The rise time of the bolometric light curve is governed by a combination of \Mni and \ejm and as noted earlier attempted to capture any uncertainties in this combination by adopting two rise times for the calculation (13 and 19 days). These rise times have been consistently applied also to the determination of \ejm. In addition, the fiducial time scale depends on the density structure of the ejecta, which may differ for the fast-declining SN\,Ia and we explored two different e-folding velocities (2700\,\kms\ and 3000\,\kms) to represent the range of possible ejecta structures. The least luminous delayed detonation models of \citet{Blondin2013} had a  density profile well characterised by an e-folding velocity of 2500\,\kms  while the typical value for more luminous models was close to 3000 \kms (which is similar to the typical e-folding velocity for the sub-\mc models of \citet{Sim2010}).  The results are presented in Table\,\ref{tab:ejm}. We take the range of results to define the uncertainty in our determination of the ejecta mass. The observational error contribution is negligible by comparison. 
 For our analysis, we use $\alpha$=1. Recent theoretical studies \citep[e.g.][]{Blondin2013,Blondin2016} find $\alpha$ to be within 20$\%$ of unity.  Even for $\alpha$ = 1.2, we would obtain a longer transparency timescale, $t_0$ by $\sim$ 15$\%$. However, the high values of $\alpha$ imply a more centrally concentrated \Nif distribution and hence, a higher value for $q$. Moreover, the high $\alpha$ values correspond to the least luminous models \citep[e.g.][]{Blondin2013}. For SNe corresponding to these luminosities, there is independent evidence for a central concentration of \Nif \citep[for e.g. from low iron line widths in nebular spectra; for e.g., see][]{Blondin2012} which would imply a significantly higher $q$ value than assumed here ($q$=1 compared to $q$=1/3 used here), hence, counterbalancing the effect of a higher $\alpha$ (see Equation~\ref{eq:ejm}) 

The derived \ejm\ of the {\em SN\,Ia-faint} are a clear indication that these are sub-Chandrasekhar explosions. The highest \ejm\ are found for the shortest rise times and the highest e-folding velocity, i.e. the shallowest density structure.

\begin{table*}
\centering
\caption{Fiducial time scales ($t_0$), ejecta masses ($M_{ej}$) and bolometric decline rate for the low-luminosity SN\,Ia with sufficient early and late time coverage to determine a peak luminosity and a late time slope (see text for assumptions about $v_e$, $\kappa$, and $q$).}
\label{tab:ejm}
\scalebox{.74}{\begin{tabular}{llccccccr}
\hline\hline
SN & $t_0$ ($t_R$ = 13\,d)  & $t_0$ ($t_R$ = 19\,d) & $M_{ej}$  ($t_R$=13\,d)  & $M_{ej}$  ( $t_R$=13\,d) & $M_{ej}$ ( $t_R$=19\,d) & $M_{ej}$ ($t_R$=19\,d) & Decline rate &\\
& \multicolumn{1}{c}{(d)}  &  \multicolumn{1}{c}{(d)} &  \multicolumn{1}{c}{(\s; $v_e$=3000 km$s^{-1}$)}     & \multicolumn{1}{c}{(\s; $v_e$=2700 km$s^{-1}$)} & \multicolumn{1}{c}{(\s; $v_e$=3000 km$s^{-1}$)} & \multicolumn{1}{c}{(\s; $v_e$=2700 km$s^{-1}$)} & \multicolumn{1}{c}{(mag d$^{-1}$)} & \\
\hline
{\em 2005ke} & 31.69 ($\pm$ 0.83)	& 28.49 ($\pm$ 0.60) & 1.03 ($\pm$  0.24) & 0.84 ($\pm$ 0.20) & 0.83 ($\pm$ 0.20) & 0.67 ($\pm$ 0.16) & 0.030 ($\pm$ 0.0004) &\\
{\em 2006mr} & 26.72 ($\pm$ 0.47)	& 24.67 ($\pm$ 0.40) & 0.73 ($\pm$ 0.19)	& 0.61 ($\pm$ 0.17) &  0.62 ($\pm$ 0.17) & 0.51 ($\pm$ 0.16) & 0.039 ($\pm$ 0.0013) &\\
{\em 2007ax}\tablefootmark{1} & 28.27 ($\pm$ 0.69) & 26.07 ($\pm$ 0.50)	& 0.81 ($\pm$ 0.21)	&  0.70 ($\pm$ 0.18) & 0.70 ($\pm$ 0.18) & 0.56 ($\pm$ 0.16) & $\ldots$  &\\
2007on & 29.70	($\pm$ 0.77) 	&	26.89 ($\pm$ 0.16) & 0.90 ($\pm$ 0.21) & 0.72 ($\pm$ 0.19) &  0.74 ($\pm$ 0.19)  & 0.60 ($\pm$ 0.16) &  0.033 ($\pm$ 0.0003) &\\
{\em 2009F}\tablefootmark{1}	&	26.49	($\pm$ 0.60)	&	24.69 ($\pm$ 0.40)	&	0.72 ($\pm$ 0.19) 		& 0.60 ($\pm$ 0.16)	& 0.62 ($\pm$ 0.16) & 0.51 ($\pm$ 0.14)	 &	$\ldots$ &	\\
\hline
\end{tabular}}
\tablefoot{\tablefoottext{1}{Only using UBVRI data}}
\end{table*}

\begin{figure}
\includegraphics[width=.5\textwidth]{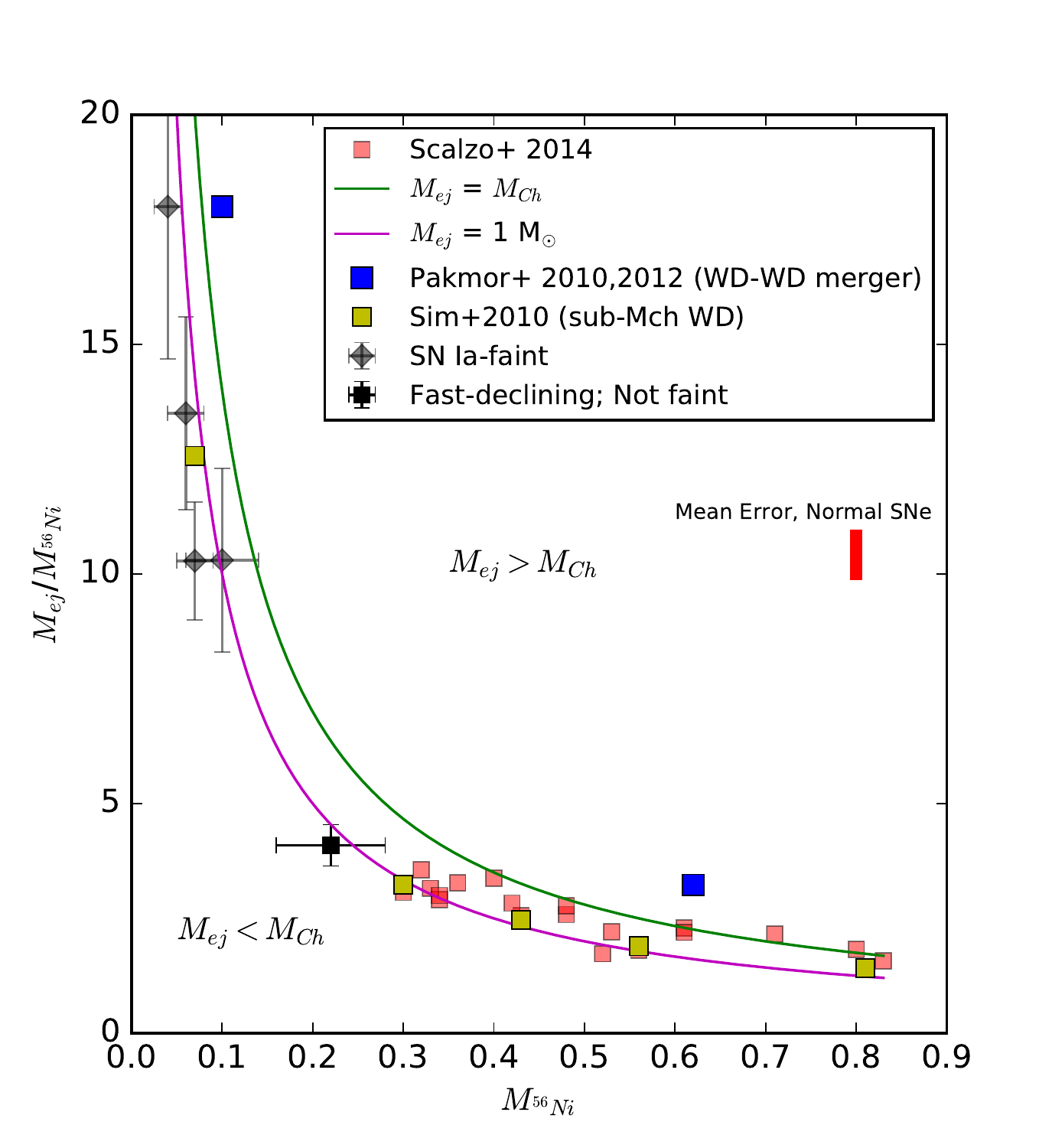}
\caption{The ratio of the ejecta mass to \Nif\ mass is plotted against the \Nif\ mass. With decreasing ejecta mass the black points are SNe\,2005ke, 2006mr, 2007on 2007ax and 2009F, the last two of which  lack NIR coverage. The diamonds are the \sfa\, subgroup and the square is 2007on which is fast-declining but not in the subgroup. The red points are normal SN\,Ia taken from \citet{Scalzo2014}. We also plot the values from different model scenarios. The yellow squares are sub-\mc double detonation models from \citet{Sim2010}, blue squares from violent merger models for normal and subluminous SNe from \citet{Pakmor2010,Pakmor2012} , the green curve is the ratio for a \mc explosion and the magenta curve is the ratio for a sub-\mc explosion with \ejm of 1 \s. We plot the typical error bar for the normal SNe from \citet{Scalzo2014} in red.}
\label{fig:ratio}
\end{figure}

Combining Tables~\ref{tab:ni} and~\ref{tab:ejm} (for a rise time of 13\,days), we calculate the ratio of the \ejm to \Mni (hereafter, $R_M$) for fast-declining SN\,Ia. Fast-declining SN~Ia in the {\em SN\,Ia-faint} sub-group have significantly larger $R_M$ values compared to normal SN\,Ia (Figure~\ref{fig:ratio}). An interesting object is SN\,2007on, which is the only fast-declining SN\,Ia connected to the normal SN\,Ia, and displays a $R_M=4.5$ and is much closer to the value of normal SN\,Ia with $R_M<4$. We also show in Figure~\ref{fig:ratio} the values from different explosion models. The observed ratios of the {\em SN\,Ia-faint} agree better with sub-\mc model values than with  the \mc values, although the errors are large due to large uncertainties in the individual \ejm and \Mni values. 

For longer rise times the derived \ejm\ decrease and the \Mni estimates increase, which leads to small $R_M$. In this case, the points tend to drop further below the line of \mc explosions. 

\section{Discussion and Conclusion}
\label{sec-disc}
The ejecta-mass estimates for our SN sample (Table~\ref{tab:ejm}) suggests that fast-declining SN~Ia are associated with sub-\mc progenitors. Pure detonations of sub-\mc WDs have also been shown to compare favourably with the narrow light curves of low-luminosity SN~Ia and hence reproduce the faint end of the width-luminosity relation \citep{Sim2010,Blondin2015b,Blondin2016}. One possible mechanism to trigger a sub-MCh WD explosion is the detonation of a surface layer of He, accreted from a companion \citep[e.g.][]{Bildsten2007}, which in turn triggers a secondary carbon detonation in the WD core \citep[known as the double detonation scenario][]{Woosley94,Livne1995,Fink2010,Shen2014}.

Two SNe in the {\em SN~Ia-faint} subgroup (SN\,1999by and SN\,2005ke) show significantly larger continuum polarisation \citep{Howell2001,Patat2012} than normal SN\,Ia \citep[e.g. see][]{Wang2007,Patat2009}. Detailed modelling of the polarisation spectra of SN\,2005ke \citep{Patat2012} led to the conclusion that it could arise from one of three scenarios; a WD rotating at close to break-up velocity, a \mc delayed detonation or a merger of two WDs. A significantly sub-\mc ejecta mass (Table~\ref{tab:ejm}) combined with
the conclusions from the polarimetry would suggest a merger of two WDs
\citep[with total mass $\mathrm{M_{tot}} < \mathrm{M_{Ch}}$;  see][]{VK2010} to be a possible scenario for SN\,2005ke. 
More spectropolarimetric observations of fast-declining SN~Ia will be key to determine their explosion mechanism.

All SNe in the {\em SN~Ia-faint} subgroup are spectroscopically 91bg-like (see Table~\ref{tab:samp}), however three SNe not in this subgroup are also classified as 91bg-like since they show strong Ti II in their maximum light spectra  \citep{Folatelli2013}. Therefore,  the presence of spectroscopic 91bg-like features is not an exclusive hallmark of the {\em SN~Ia-faint} subgroup, although the Ti II feature in the  \sfa\, subgroup SNe is stronger than the feature in the three SNe not in this subgroup.

The {\em SN~Ia faint} subgroup are characterised by single-peaked NIR light curves. \citet{Kasen2006} find in their lowest \Nif\, mass models that the NIR light curves are single peaked since the shift from doubly to singly ionized iron group elements (IGEs) that creates the second maximum occurs only  $\sim$ 20\,days after explosion hence, coinciding with the primary maximum. Objects in our {\em SN\,Ia-faint} subgroup have inferred \Nif masses $\lesssim$ 0.1 \s indicating that their different NIR light curve morphology is a direct result of the low \Nif\, yield. We note that other fast-declining SNe, e.g. 1991bg, 1998de, show no $i$-band second maximum \citep{Filippenko1992b,Turatto1996,Modjaz2001} and a late $i$-band peak (similar to the NIR properties of the \sfa\, subgroup) which would also make them members of this subclass.

Type Iax supernovae \citep{Foley2013} also show a single NIR maximum despite displaying a large range of decline rates (1.2 \textless \dm \textless 2.4) and inferred \Nif\, masses ($\sim$ 0.001 - 0.18 M$_{\odot}$). This is understood as a result of a high degree of mixing of \Nif\, \citep[seen in 3-D deflagration models; e.g.][]{Kromer2013,Fink2014} in the ejected material and is observationally supported by the rapid rise of the light curve to maximum \citep[see for e.g.][]{Yamanaka2015} and the presence of iron in early time spectra \citep[for e.g.][]{Li2003}.. The nebular phase spectra of SN~Iax, which, in some cases show P-Cygni line profiles, unlike the forbidden iron lines in nebular spectra of SN~Ia \citep[ e.g.][]{Jha2006,Foley2016}, along with the peculiar maximum light spectroscopic and photometric properties would point towards them being distinct explosions from the fast-declining SNe analysed here.

Pure deflagrations which only partially unbind the progenitor, leaving a bound remnant \citep[e.g.][]{Kromer2013,Fink2014} can explain the low \ejm\, of the \sfa\, subgroup, though the corresponding \Nif\, masses are higher than the values inferred from observations (Table~\ref{tab:ni}). Pure deflagration models with a significantly lower \ejm\, ($\sim$ 0.2 \s) agree well with the $B$-band and bolometric light curves of \sfa\, subgroup member SN~2005bl \citep{Tauben2008,Fink2014}, but cannot explain the extremely red colours for these SNe. We note, however, this explosion mechanism would not be a viable candidate for fast-decliners with two NIR maxima (e.g. SN~2007on) since the \Nif is highly mixed in the ejecta, producing theoretical light curves with only a single maximum in the NIR.

We have shown that SN\,Ia that exhibit \sbv below 0.5  deviate from the \lm-\sbv relation for normal SN~Ia, show a low NIR peak luminosity, late NIR maxima and also lack a prominent second maximum in the NIR filters. This behaviour distinguishes them from fast-declining SN\,Ia with \sbv above 0.5 which seem to extend the normal SN\,Ia sequence to fainter magnitudes. From this work it is evident that the \sbv metric of \citet{Burns2014} is a powerful diagnostic of the nature of the explosion. 
From the low bolometric luminosity we infer small \Nif\, mass and from fitting an energy deposition function to the tail of the bolometric light curve, we infer a sub-\mc ejecta mass. The low values for these global parameters, combined with the differences in the NIR and bolometric properties of the two subgroups of fast-declining SN~Ia could point to two different explosion scenarios leading to fast-declining SN~Ia.



\nocite{*}
\begin{thebibliography}{99}
\bibliographystyle{aa}

\bibitem[\protect\citeauthoryear{Ajhar et al.}{2001}]{Ajhar2001} 
Ajhar E.~A., Tonry J.~L., Blakeslee J.~P., Riess A.~G., Schmidt B.~P., 
2001, ApJ, 559, 584 




\bibitem[\protect\citeauthoryear{Arnett}{1982}]{Arnett1982} Arnett 
W.~D., 1982, ApJ, 253, 785 

\bibitem[\protect\citeauthoryear{Arnett et al.}{1985}]{Arnett1985} Arnett 
W.~D., Branch, D., Wheeler, J. C., 1985, Naturce, 314, 337





\bibitem[\protect\citeauthoryear{Benetti et 
al.}{2004}]{Benetti2004} Benetti S., et al., 2004, MNRAS, 348, 261

\bibitem[Bildsten et al.(2007)]{Bildsten2007} Bildsten, L., Shen, 
K.~J., Weinberg, N.~N., \& Nelemans, G.\ 2007, \apjl, 662, L95 


\bibitem[\protect\citeauthoryear{Blondin et 
al.}{2012}]{Blondin2012} Blondin S., et al., 2012, AJ, 143, 126 

\bibitem[\protect\citeauthoryear{Blondin et 
al.}{2013}]{Blondin2013} Blondin S., Dessart L., Hillier D.~J., 
Khokhlov A.~M., 2013, MNRAS, 429, 2127 

\bibitem[\protect\citeauthoryear{Blondin, Dessart, 
\& Hillier}{2015}]{Blondin2015} Blondin S., Dessart L., Hillier D.~J., 2015, MNRAS, 448, 2766 

\bibitem[\protect\citeauthoryear{Blondin et al.}{2016}]{Blondin2016} Blondin S., et al., 2016, MNRAS, submitted 

\bibitem[\protect\citeauthoryear{Blondin}{2015}]{Blondin2015b} 
Blondin S., 2015, sf2a.conf, 319 

\bibitem[\protect\citeauthoryear{Branch}{1992}]{Branch1992} Branch 
D., 1992, ApJ, 392, 35

\bibitem[\protect\citeauthoryear{Burns et al.}{2011}]{Burns2011} 
Burns C.~R., et al., 2011, AJ, 141, 19 

\bibitem[\protect\citeauthoryear{Burns et al.}{2014}]{Burns2014} Burns
C.~R., et al., 2014, ApJ, 789, 32 







\bibitem[\protect\citeauthoryear{Contardo, Leibundgut 
\& Vacca}{2000}]{Contardo2000} Contardo G., Leibundgut B., Vacca W.~D., 2000, A\&A, 359, 876 

\bibitem[\protect\citeauthoryear{Contreras et 
al.}{2010}]{Contreras2010} Contreras C., et al., 2010, AJ, 139, 519 

\bibitem[\protect\citeauthoryear{Dhawan et al.}{2015}]{Dhawan2015} 
Dhawan S., Leibundgut B., Spyromilio J., Maguire K., 2015, MNRAS, 448, 1345

\bibitem[Dhawan et al.(2016)]{Dhawan2015b} Dhawan, S., Leibundgut, B., Spyromilio, J., \& Blondin, S.\ 2016, \aap, 588, A84 








\bibitem[\protect\citeauthoryear{Elias et al.}{1981}]{Elias1981} Elias, J.~H., Frogel, J.~A., Hackwell, J.~A., Persson, E.~E., 1981, ApJ, 251, L13



\bibitem[\protect\citeauthoryear{Feroz et al.}{2013}]{Feroz2013} Feroz F., Hobson M.~P., Cameron E., Pettitt A.~N., 2013, arXiv, arXiv:1306.2144 


\bibitem[\protect\citeauthoryear{Filippenko et 
al.}{1992}]{Filippenko1992a} Filippenko A.~V., et al., 1992, AJ, 104, 
1543 


\bibitem[\protect\citeauthoryear{Filippenko et 
al.}{1992}]{Filippenko1992b} Filippenko A.~V., et al., 1992, ApJ, 384, 
L15 

\bibitem[\protect\citeauthoryear{Fink et 
al.}{2010}]{Fink2010} Fink M., R{\"o}pke F.~K., Hillebrandt W., Seitenzahl I.~R., Sim S.~A., Kromer M., 2010, A\&A, 514, A53 

\bibitem[\protect\citeauthoryear{Fink et al.}{2014}]{Fink2014} Fink M., et al., 2014, MNRAS, 438, 1762 



\bibitem[\protect\citeauthoryear{Folatelli et 
al.}{2010}]{Folatelli2010} Folatelli G., et al., 2010, AJ, 139, 120 

\bibitem[\protect\citeauthoryear{Folatelli et 
al.}{2013}]{Folatelli2013} Folatelli G., et al., 2013, ApJ, 773, 53 

\bibitem[\protect\citeauthoryear{Foley et al.}{2009}]{Foley2009} Foley, R.~J., et al., AJ, 138, 376




\bibitem[\protect\citeauthoryear{Fransson \& Jerkstrand}{2015}] {Fransson2015} Fransson, C., Jerkstrand, A., 2015, ApJ, 814, L2

\bibitem[\protect\citeauthoryear{Freedman et 
al.}{2001}]{Freedman2001} Freedman W.~L., et al., 2001, ApJ, 553, 47 


\bibitem[\protect\citeauthoryear{Friedman et 
al.}{2015}]{Friedman2014} Friedman A.~S., et al., 2015, ApJS, 220, 9 


\bibitem[\protect\citeauthoryear{Foley et al.}{2013}]{Foley2013} Foley R.~J., et al., 2013, ApJ, 767, 57 
\bibitem[\protect\citeauthoryear{Foley et al.}{2016}]{Foley2016} Foley R.~J., Jha S.~W., Pan Y.-C., Zheng W.~K., Bildsten L., Filippenko A.~V., Kasen D., 2016, MNRAS, 461, 433 


\bibitem[\protect\citeauthoryear{Garnavich et 
al.}{2004}]{Garnavich2004} Garnavich P.~M., et al., 2004, ApJ, 613, 
1120 












\bibitem[\protect\citeauthoryear{Hillebrandt 
\& Niemeyer}{2000}]{Hillebrandt2000} Hillebrandt W., Niemeyer J.~C., 2000, ARA\&A, 38, 191 







\bibitem[\protect\citeauthoryear{H{\"o}flich et 
al.}{2002}]{Hoeflich2002} H{\"o}flich P., Gerardy C.~L., Fesen 
R.~A., Sakai S., 2002, ApJ, 568, 791 



\bibitem[\protect\citeauthoryear{Howell}{2001}]{Howell2001} Howell 
D.~A., 2001, ApJ, 554, L193 

\bibitem[Howell et al.(2001)]{Howell2001a} Howell, D.~A., 
H{\"o}flich, P., Wang, L., \& Wheeler, J.~C.\ 2001, \apj, 556, 302 


\bibitem[\protect\citeauthoryear{Hoyle 
\& Fowler}{1960}]{Hoyle1960} Hoyle F., Fowler W.~A., 1960, ApJ, 132, 565 


\bibitem[\protect\citeauthoryear{Hsiao et 
al.}{2015}]{Hsiao2015} Hsiao E.~Y., et al., 2015, A\&A, 578, A9 




\bibitem[\protect\citeauthoryear{Jeffery}{1999}]{Jeffrey1999} 
Jeffery D.~J., 1999, astro, arXiv:astro-ph/9907015 
 


\bibitem[\protect\citeauthoryear{Jensen et
al.}{2003}]{Jensen2003} Jensen J.~B., Tonry J.~L., Barris
B.~J., Thompson R.~I., Liu M.~C., Rieke M.~J., Ajhar E.~A., Blakeslee
J.~P., 2003, ApJ, 583, 712 

\bibitem[\protect\citeauthoryear{Jha et al.}{2006}]{Jha2006} Jha S., Branch D., Chornock R., Foley R.~J., Li W., Swift B.~J., Casebeer D., Filippenko A.~V., 2006, AJ, 132, 189 

\bibitem[\protect\citeauthoryear{Kasen}{2006}]{Kasen2006} Kasen 
D., 2006, ApJ, 649, 939

 
 






\bibitem[\protect\citeauthoryear{Kattner et al.}{2012}]{Kattner2012} Kattner S., et al., 2012, PASP, 124, 114 






\bibitem[\protect\citeauthoryear{Kromer et al.}{2013}]{Kromer2013} Kromer M., et al., 2013, MNRAS, 429, 2287 


\bibitem[\protect\citeauthoryear{Kromer et al.}{2015}]{Kromer2015} Kromer, M. et al., 2015, MNRAS, 450, 3035





\bibitem[\protect\citeauthoryear{Krisciunas et al.}{2004}]{Krisciunas2004} Krisciunas K., Phillips M.M., Suntzeff, N.B., 2004, ApJ, 602, 81




\bibitem[\protect\citeauthoryear{Krisciunas et 
al.}{2009}]{Krisciunas2009} Krisciunas K., et al., 2009, AJ, 138, 1584 




 



\bibitem[\protect\citeauthoryear{Leibundgut et 
al.}{1993}]{Leibundgut1993} Leibundgut B., et al., 1993, AJ, 105,
301 





\bibitem[\protect\citeauthoryear{Li et al.}{2003}]{Li2003} Li W., et al., 2003, PASP, 115, 453 

\bibitem[\protect\citeauthoryear{Li et al.}{2011}]{Li2011} Li 
W., et al., 2011, MNRAS, 412, 1441 

\bibitem[\protect\citeauthoryear{Livne 
\& Arnett}{1995}]{Livne1995} Livne E., Arnett D., 1995, ApJ, 452, 62 







\bibitem[\protect\citeauthoryear{Mazzali et 
al.}{1997}]{Mazzali1997} Mazzali P.~A., Chugai N., Turatto M., Lucy 
L.~B., Danziger I.~J., Cappellaro E., della Valle M., Benetti S., 1997, 
MNRAS, 284, 151 




\bibitem[\protect\citeauthoryear{Meikle}{2000}]{Meikle2000} Meikle 
W.~P.~S., 2000, MNRAS, 314, 782 

\bibitem[\protect\citeauthoryear{Modjaz et al.}{2001}]{Modjaz2001} 
Modjaz M., Li W., Filippenko A.~V., King J.~Y., Leonard D.~C., Matheson T., 
Treffers R.~R., Riess A.~G., 2001, PASP, 113, 308 







\bibitem[\protect\citeauthoryear{Pakmor et al.}{2010}]{Pakmor2010} 
Pakmor R., Kromer M., R{\"o}pke F.~K., Sim S.~A., Ruiter A.~J., Hillebrandt 
W., 2010, Nature, 463, 61 



\bibitem[\protect\citeauthoryear{Pakmor et al.}{2012}]{Pakmor2012} 
Pakmor R., Kromer M., Taubenberger S., et al., 2012, ApJ, 747, L10 



\bibitem[Patat et 
al.(2009)]{Patat2009} Patat, F., Baade, D., H{\"o}flich, P., et al.\ 2009, \aap, 508, 229 


\bibitem[Patat et 
al.(2012)]{Patat2012} Patat, F., H{\"o}flich, P., Baade, D., et al.\ 2012, \aap, 545, A7 











\bibitem[\protect\citeauthoryear{Phillips}{1993}]{Phillips1993} 
Phillips M.~M., 1993, ApJ, 413, L105 

\bibitem[\protect\citeauthoryear{Phillips}{2012}]{Phillips2012} 
Phillips M.~M., 2012, PASA, 29, 434 


\bibitem[\protect\citeauthoryear{Phillips et 
al.}{1992}]{Phillips1992} Phillips M.~M., Wells L.~A., Suntzeff 
N.~B., Hamuy M., Leibundgut B., Kirshner R.~P., Foltz C.~B., 1992, AJ, 103, 
1632 
%

\bibitem[\protect\citeauthoryear{Phillips et 
al.}{1999}]{Phillips1999} Phillips M.~M., Lira P., Suntzeff N.~B., 
Schommer R.~A., Hamuy M., Maza J., 1999, AJ, 118, 1766 


\bibitem[\protect\citeauthoryear{Phillips et al.}{2007}] {Phillips2007} Phillips, M.~M., et al., 2007, PASP, 119, 360



\bibitem[\protect\citeauthoryear{Pinto 
\& Eastman}{2000}]{Pinto2000} Pinto P.~A., Eastman R.~G., 2000, ApJ, 530, 757 


















\bibitem[\protect\citeauthoryear{Scalzo et al.}{2014}]{Scalzo2014} 
Scalzo R., et al., 2014, MNRAS, 440, 1498 



\bibitem[\protect\citeauthoryear{Shen \& Moore}{2014}]{Shen2014} Shen K.~J., Moore K., 2014, ApJ, 797, 46 




\bibitem[\protect\citeauthoryear{Sim et al.}{2010}]{Sim2010} 
Sim S.~A., et al., 2010, ApJ, 714, L52 
Ruiter A.~J., Seitenzahl I.~R.,


\bibitem[\protect\citeauthoryear{Skilling}{2004}]{Skilling2004} Skilling J., 2004, AIPC, 735, 395 



\bibitem[\protect\citeauthoryear{Stritzinger et
al.}{2006}]{Stritzinger2006a} Stritzinger M., Leibundgut B., Walch S.,
Contardo G., 2006, A\&A, 450, 241 




\bibitem[\protect\citeauthoryear{Stritzinger et 
al.}{2011}]{Stritzinger2011} Stritzinger M.~D., et al., 2011, AJ, 142, 
156 

\bibitem[\protect\citeauthoryear{Swartz, Sutherland, 
\& Harkness}{1995}]{Swartz1995} Swartz D.~A., Sutherland P.~G., Harkness R.~P., 1995, ApJ, 446, 766 




\bibitem[\protect\citeauthoryear{Taubenberger et 
al.}{2008}]{Tauben2008} Taubenberger S., et al., 2008, MNRAS, 385, 
75 



\bibitem[\protect\citeauthoryear{Tomasella et al.}{2016}] {Tomasella2016} Tomasella, L., et al., MNRAS, 459, 1018
\bibitem[\protect\citeauthoryear{Turatto et al.}{1996}]{Turatto1996} Turatto M., Benetti S., Cappellaro E., Danziger I.~J., Della Valle M., Gouiffes C., Mazzali P.~A., Patat F., 1996, MNRAS, 283, 1 

\bibitem[\protect\citeauthoryear{Tonry et al.}{2001}]{Tonry2001} 
Tonry J.~L., Dressler A., Blakeslee J.~P., Ajhar E.~A., Fletcher A.~B., 
Luppino G.~A., Metzger M.~R., Moore C.~B., 2001, ApJ, 546, 681


\bibitem[\protect\citeauthoryear{Trotta}{2008}]{Trotta2008} Trotta R., 2008, ConPh, 49, 71 

\bibitem[\protect\citeauthoryear{Tully et al.}{2009}]{Tully2009} Tully R.~B., Rizzi L., Shaya E.~J., Courtois H.~M., Makarov D.~I., Jacobs B.~A., 2009, AJ, 138, 323 
\bibitem[\protect\citeauthoryear{Tully et al.}{2013}]{Tully2013} Tully R.~B., et al., 2013, AJ, 146, 86 





\bibitem[\protect\citeauthoryear{van Kerkwijk, Chang, 
\& Justham}{2010}]{VK2010} van Kerkwijk M.~H., Chang P., Justham S., 2010, ApJ, 722, L157 

\bibitem[Wang et al.(2007)]{Wang2007} Wang, L., Baade, D., 
\& Patat, F.\ 2007, Science, 315, 212 



\bibitem[\protect\citeauthoryear{Wood-Vasey et 
al.}{2008}]{Wood-Vasey2008} Wood-Vasey W.~M., et al., 2008, ApJ, 689, 
377 


 

\bibitem[\protect\citeauthoryear{Woosley 
\& Weaver}{1994}]{Woosley94} Woosley S.~E., Weaver T.~A., 1994, ApJ, 423, 371 



\bibitem[\protect\citeauthoryear{Yamanaka et al.}{2015}]{Yamanaka2015} Yamanaka M., et al., 2015, ApJ, 806, 191 


\end{thebibliography}


\begin{acknowledgements}
This research was supported by the DFG Cluster of Excellence ʻOrigin and
Structure of the Universe'.
B.L. acknowledges support for this work by the Deutsche
Forschungsgemeinschaft through TRR33, The Dark Universe. We all are grateful to the ESO Visitor Programme to support the visit of S. B. to Garching when this work was started.
We  thank Andrew Friedman for providing CfAIR2 light curves in machine-readable form and Stefan Taubenberger for discussions on rise times of fast declining SN~Ia.

\end{acknowledgements}
\end{document}